\documentstyle[12pt]{article}
\def\be{\begin{equation}}
\def\ee{\end{equation}}
\def\bea{\begin{eqnarray}}
\def\eea{\end{eqnarray}}

\begin{document}

\begin{center}
{\Large{\bf Some $\mathbf{Z}_2$-invariant Variables:\\
Constructed of the Scalars Dilaton and Axion}}

\vskip .5cm
{\large Davoud Kamani}
\vskip .1cm
{\it Department of Physics, Amirkabir University of Technology 
(Tehran Polytechnic)\\  
P.O.Box: 15875-4413, Tehran, Iran}\\
{\sl e-mail: kamani@aut.ac.ir}\\
\end{center}

\begin{abstract}

Using the dilaton scalar and axion pseudoscalar fields 
we construct a number of scalars and differential 
forms which are symmetric under the $\mathbf{Z}_2$-subgroup 
of the group $SL(2, \mathbf{R})$. These invariants enable 
us to  establish various 10-dimensional invariant actions.
Other invariants which are not independent from the previous 
ones will be detached.  

\end{abstract}

{\it PACS numbers}: 11.25.-w

{\it Keywords}: Dilaton; axion; invariant variables.

\vskip .5cm
\newpage

\section{Introduction}

Deformation and 
generalization of the supergravity actions have been studied from
the various point of view. Among them the modified versions of the 
type IIB supergravity are more desirable. One reason is due to 
the fact that the type IIB theory contains 
the axion (the R-R pseudoscalar field) and dilaton 
(the NS-NS scalar field), which are corresponding 
to the D-instanton and the Hodge duals of their field strengths are
associated with the D7-brane and NS7-brane. 
The role of the type IIB instanton has been extremely studied in
the subject of the AdS/CFT correspondence.
In addition, the axion of the type IIB theory solves some 
cosmological problems \cite{1,2,3}.

On the other hand there is the $SL(2, \mathbf{R})$ group which is 
certainly a symmetry group of the type IIB supergravity. 
There is also an
important special case, i.e. the S-duality, which relates strong 
and weak coupling phases of a given theory in some cases, whereas 
in some other situations strong and weak coupling regimes of two 
different theories are connected. 
 
In this paper, from the dilaton and axion fields and their field 
strengths and also the Hodge duals of their field strengths we 
construct various differential $r$-forms with 
$r=0, 1, 2, 10$ which are invariant under the 
$\mathbf{Z}_2$-subgroup of the group $SL(2, \mathbf{R})$. 
These differential forms enable us to obtain 
various independent 10-dimensional actions which separately 
have the above symmetry. In fact, many other invariant 
scalars, differential forms and actions can be made which 
are not independent of the previous ones. We shall  
detach independent invariants.  

The following facts motivated us to study the $\mathbf{Z}_2$-invariant
variables specially the invariant actions. The first is that,
since the Einstein-Hilbert action also is a $\mathbf{Z}_2$-invariant, 
it is relevant to add a subset of the invariant actions to this 
action. In this case, the full resulted theory possesses this symmetry. 
Thus, this subset of the actions becomes a source for the gravity. 
In addition, appearance of the 1-forms and 9-forms are 
appropriate for realizing the D-instanton \cite{4}, D7-branes and NS7-branes.
Furthermore, investigation on the dilaton and axion clarifies 
some properties of the D7-brane and NS7-brane, and hence may 
shed light on the F-theory \cite{5}. 
Beside, we shall observe that the $\mathbf{Z}_2$-transformations 
can be viewed as gauge transformations. Finally, since the 
exponential of the dilaton defines the string coupling 
we should know all possible dynamics of this field, which 
are associated with various actions.

This paper is organized as follows. In Sec. 2, the 
$\mathbf{Z}_2$-invariant scalars will be constructed. In Sec. 3, 
the $\mathbf{Z}_2$-invariant differential forms will be made. 
In Sec. 4, by using the above scalars and differential 
forms, various $\mathbf{Z}_2$-invariant actions will be established. 
In Sec. 5, independent invariants will be detached 
from the dependent ones. Section 6 is devoted to the conclusions.
\section{Invariant scalars}

We remember that the bosonic massless excitations of the type IIB 
theory includes the graviton $G_{\mu\nu}$, dilaton $\Phi$ and 
antisymmetric tensor $B_{\mu\nu}$ in the NS-NS sector, and the 
R-R counterparts are the axion $C_0 \equiv C$, an antisymmetric tensor 
field $C_{(2)\mu\nu}$ and a four-index antisymmetric potential 
$C_{(4)\mu\nu\rho\lambda}$ which its field strength is self-dual.
In addition, in both sectors the Hodge duality also procreates
some other form fields.

The $\mathbf{Z}_2$-duality is generated by the $\mathbf{Z}_2$-subgroup 
of the group $SL(2, \mathbf{R})$. 
This determines $\mathbf{Z}_2$-transformations of the scalar fields 
$C$ and $\Phi$ as $\tau \rightarrow \tau'=
-1/\tau$ where $\tau=C + ie^{-\Phi}$ is the complex 
axion-dilaton modulus. This transformation mixes the two scalars 
as in the following
\bea
&~& e^{-\Phi'}= \frac{e^{-\Phi}}{C^2+ e^{-2\Phi}}\;,
\nonumber\\
&~& C'= -\frac{C}{C^2+ e^{-2\Phi}}\;.
\eea
Since the string coupling is given by $g_s=e^{\Phi}$, when the axion 
vanishes, the first equation relates the weak and strong coupling regimes 
of the type IIB superstring theory. The Einstein metric, which 
will be used, is an $\mathbf{Z}_2$-invariant variable
\bea
G'_{{\rm E}\mu\nu}=G_{{\rm E}\mu\nu},
\eea
where the Einstein frame and string frame metrics are related by
$G_{{\rm E}\mu\nu}=e^{-\Phi/2}G_{\mu\nu}$.
From now on we utilize the word ``{\it invariant}'' 
instead of ``{\it $\mathbf{Z}_2$-invariant}'', 
which means invariance under the transformations (1) and (2).

According to the Eqs. (1) we acquire the following equations 
\bea
&~& \frac{e^{-2\Phi'}}{C'^2 + e^{-2\Phi'}}=\frac{e^{-2\Phi}}
{C^2 + e^{-2\Phi}} \equiv \sigma_1 (C , \Phi)\;,
\nonumber\\
&~& \frac{C'^2}{C'^2 + e^{-2\Phi'}}=\frac{C^2}{C^2 
+ e^{-2\Phi}} \equiv \sigma_2 (C , \Phi)\;.
\eea
These equations represent two invariant scalars $\sigma_1$
and $\sigma_2$, i.e. $\sigma_i (C' , \Phi')=\sigma_i (C , \Phi)$
for $i=1, 2$. 

Besides, there are also the following invariant scalars 
which comprise derivatives of the fields $\Phi$ and $C$,
\bea
e^{2\Phi'}F'_\mu F'^\mu + H'_\mu H'^\mu=
e^{2\Phi}F_\mu F^\mu + H_\mu H^\mu,
\eea
\bea 
e^{2\Phi'}[2C' F'_\mu H'^\mu +(C'^2 - e^{-2\Phi'}) H'_\mu H'^\mu] =
e^{2\Phi}[2C F_\mu H^\mu +(C^2 - e^{-2\Phi}) H_\mu H^\mu],
\eea
\bea
e^{4\Phi'}[(C'^2 - e^{-2\Phi'}) F'_\mu F'^\mu 
-2C' e^{-2\Phi'} F'_\mu H'^\mu] =
e^{4\Phi}[(C^2 - e^{-2\Phi}) F_\mu F^\mu 
-2C e^{-2\Phi} F_\mu H^\mu],
\eea
where $H_\mu=\partial_\mu \Phi$, $F_\mu=\partial_\mu C$,
$H'_\mu=\partial_\mu \Phi'$ and $F'_\mu=\partial_\mu C'$.
The indices are raised by the Einstein metric, e.g.  
$F^\mu = G^{\mu\nu}_{\rm E}F_\nu$. Some other kinds of the 
invariant scalars can be seen in the integrand of the action 
(24) and in the Eq. (27), where, in fact, the latter is 
anti-invariant. 

Note that the 1-form field strength $F=F_\mu dx^\mu$ is
corresponding to the D(-1)-brane (D-instanton). This brane 
is unique among the D-branes since it is localized in time as 
well as in space.   
\section{Invariant differential forms}

Exterior derivatives of the scalar fields give 1-forms. Combination 
of these 1-forms and their Hodge duals determines some higher order 
invariant differential forms which will be used to obtain several 
invariant actions.

$\bullet$ {\bf Invariant 2-form}

Exterior derivative of the Eqs. (1) defines the 1-forms
\bea
&~& H' = \frac{1}{C^2 + e^{-2\Phi}}\bigg{(}2C F+(C^2
- e^{-2\Phi})H \bigg{)},
\nonumber\\
&~& F'=\frac{1}{(C^2 + e^{-2\Phi})^2}\bigg{(}(C^2 
- e^{-2\Phi})F
-2C e^{-2\Phi}H \bigg{)}.
\eea 
Wedge product of these forms specifies the following invariant 
2-form 
\bea
e^{\Phi'} F' \wedge H'=e^{\Phi} F \wedge H \equiv {\cal{F}}.
\eea
This equation elaborates that ${\cal{F}}$ is field strength of the 1-form 
$A=C e^\Phi H$, and or $A'=C' e^{\Phi'} H'$. These
1-forms are related to each other through the gauge transformation 
$A'=A+\Lambda$ where $\Lambda$ is a closed 1-form. 
We shall investigate this.

$\bullet$ {\bf Invariant 10-forms}

We know that Hodge duality definition of a differential form is
characterized by the metric of the manifold. According to this, 
for the next purposes, we define this duality via the 
Einstein metric. For example, the Hodge dual of the 1-form 
$F$, which is a 9-form ${\widetilde F}$, has the components
\bea
{\widetilde F}_{\mu_1\mu_2\cdot\cdot\cdot \mu_9}=
\sqrt{-G_{\rm E}}F_\mu G_{\rm E}^{\mu\nu}
\varepsilon_{\nu\mu_1\mu_2\cdot\cdot\cdot \mu_9},
\eea
where $G_{\rm E} =\det G_{{\rm E}\mu\nu}$. The Levi-Civita symbol 
$\varepsilon_{\mu_1\mu_2\cdot\cdot\cdot \mu_{10}}$ has the 
components $\pm 1$ and 0. Therefore, the Hodge duality on the 
Eqs. (7) exhibits the following 9-forms
\bea
&~& {\tilde H'} = \frac{1}{C^2 
+ e^{-2\Phi}}\bigg{(}2C {\widetilde F}
+(C^2- e^{-2\Phi}){\tilde H} \bigg{)},
\nonumber\\
&~& {\widetilde F'}=\frac{1}{(C^2 + e^{-2\Phi})^2}\bigg{(}
(C^2 - e^{-2\Phi}){\widetilde F}
-2C e^{-2\Phi}{\tilde H} \bigg{)}.
\eea

Let us give a brief description regarding the 9-forms ${\widetilde F}$ 
and ${\tilde H}$. Two different local combinations of these forms 
accompanied by other form fields
have been associated with the D7-brane and NS7-brane. The 
NS7-brane is related to the D7-brane by the S-duality. 
These transformations also imply 
that the D7-brane and NS7-brane do not form a doublet under 
the S-duality. It has been commonly accepted that there are bound 
states of $p$ D7-branes and $q$ NS7-branes which transform as 
doublets. The D7-brane has a magnetic charge whereas its dual 
is the D-instanton, which has an electric charge. 

Combining (7) with (10) 
via the wedge product specifies the following invariant 10-forms
\bea
e^{2\Phi'}F' \wedge {\widetilde F'}+H'\wedge {\tilde H'}
=e^{2\Phi}F \wedge {\widetilde F}+H\wedge {\tilde H}.
\eea
\bea
e^{2\Phi'}[2C'  F' \wedge {\tilde H'}+(C'^2-e^{-2\Phi'})H'\wedge {\tilde H'}]
=e^{2\Phi}[2C  F \wedge {\tilde H}+(C^2-e^{-2\Phi})H\wedge {\tilde H}],
\eea
\bea
e^{4\Phi'}[(C'^2-e^{-2\Phi'})F' \wedge {\widetilde F'}-2C' e^{-2\Phi'}
H' \wedge {\widetilde F'}]=e^{4\Phi}[(C^2-e^{-2\Phi})F 
\wedge {\widetilde F}-2C e^{-2\Phi}
H \wedge {\widetilde F}].
\eea
These 10-forms have been precisely written on the basis 
of the Einstein frame. Note that there is also another invariant 10-form 
$e^\Phi ( H \wedge {\widetilde F} - F \wedge {\tilde H})$ 
which identically vanishes.
\section{Invariant actions}

Making use of the 10-forms (11) - (13) we establish 
the following actions, which have the $\mathbf{Z}_2$-symmetry
\bea
I_1 = -\frac{1}{9! 4\kappa^2} \int_{\cal{M}} 
(F \wedge {\widetilde F_{\rm s}}+e^{-2\Phi}H\wedge {\tilde H_{\rm s}}),
\eea
\bea
I_2 = -\frac{1}{9! 4\kappa^2} \int_{\cal{M}} 
[2C  F \wedge {\tilde H}_{\rm s}+(C^2-e^{-2\Phi})H\wedge {\tilde H}_{\rm s}],
\eea
\bea
I_3 = -\frac{1}{9! 4\kappa^2} \int_{\cal{M}} 
e^{2\Phi}[(C^2-e^{-2\Phi})F \wedge {\widetilde F}_{\rm s}-2C e^{-2\Phi}
H \wedge {\widetilde F}_{\rm s}],
\eea
where $\kappa$ is the 10-dimensional gravitational constant 
and ${\cal{M}}$ indicates the spacetime manifold. These actions have 
been written in the string frame which is more usual. 
The 9-forms in the two frames are related by   
${\widetilde F}=e^{-2\Phi}{\widetilde F_{\rm s}}$ and 
${\tilde H}=e^{-2\Phi}{\tilde H_{\rm s}}$ where  
the subscript ``s'' refers to the string frame.

The other forms of the above actions can be made by the standard 
duality transformation 
\bea
S_1 = -\frac{1}{4\kappa^2} \int_{\cal{M}} d^{10} x \sqrt{-G}
(F_\mu F^\mu_{\rm s} + e^{-2\Phi} H_\mu H^\mu_{\rm s}),
\eea
\bea
S_2 = -\frac{1}{4\kappa^2} \int_{\cal{M}} d^{10} x \sqrt{-G}
[2C F_\mu H^\mu_{\rm s} +(C^2 - e^{-2\Phi}) H_\mu H^\mu_{\rm s}],
\eea
\bea
S_3 = -\frac{1}{4\kappa^2} \int_{\cal{M}} d^{10} x \sqrt{-G}
e^{2\Phi}[(C^2 - e^{-2\Phi}) F_\mu F^\mu_{\rm s} 
-2C e^{-2\Phi} F_\mu H^\mu_{\rm s} ],
\eea
where $G=\det G_{\mu\nu}$, and $F^\mu_{\rm s}$ and 
$H^\mu_{\rm s}$ refer to the string frame, i.e. 
$F^\mu_{\rm s}=G^{\mu\nu}F_\nu$ and $H^\mu_{\rm s}=G^{\mu\nu}H_\nu$.
According to the invariant scalars (4) - (6) these actions also 
are symmetric under the $\mathbf{Z}_2$-transformations.
The action $S_1$ in the Einstein frame can be written in the form  
\bea
S_1 = -\frac{1}{4\kappa^2} \int_{\cal{M}} d^{10} x \sqrt{-G_{\rm E}}
\;\frac{\partial_\mu \tau\partial^\mu {\bar \tau}}{({\rm Im}\tau)^2}.
\eea
This clarifies that $S_1$ is a part of the action of the type IIB 
supergravity, which we found an alternative derivation 
and various feature for it. Note that the actions $S_1$, $S_2$ and 
$S_3$ are not independent of the actions $I_1$, $I_2$ and $I_3$. 
In other words, the action $S_i$ is an alternative version of the 
action $I_i$ for $i \in \{1, 2, 3\}$. 

The effective field theory actions describing the dynamics of 
the massless modes of the various string theories contain, in addition 
to the well-known supergravity terms, an infinite number of 
higher-derivative corrections. Symmetries and dualities of these 
effective field theories are useful tools to find such higher-derivative 
terms. Therefore, several explicit higher-derivative terms have been 
known or conjectured to exist by symmetry/duality arguments 
(e.g. see \cite{6} and references therein). Specially, the effective 
action for bosonic massless modes of the type IIB supergravity 
at various levels of derivatives has been studied. Naturally, 
the dilaton and axion fields also contribute on higher order 
derivative corrections to the type IIB theory \cite{7, 8, 9}. 
Though the proposed forms of the corrections to the effective 
action satisfy the $SL(2,R)$-invariance condition, but there 
is no proof for their validity \cite{8}. On the other hand, 
string scattering diagrams describe a large number of the 
Feynman diagrams in the Yang-Mills gauge theory \cite{10}. 
It has been realized that only the bosonic degrees of freedom 
are relevant in the field theoretical limit of string 
amplitudes \cite{11}. Thus, according to these facts, we 
proceed to construct a four-derivative action.

The 2-form (8) has the components
\bea
{\cal{F}}_{\mu\nu}= e^{\Phi}(F_\mu H_\nu - F_\nu H_\mu).
\eea
This indicates a field strength   
${\cal{F}}_{\mu\nu}= \partial_\mu A_\nu-\partial_\nu A_\mu$
where the $U(1)$ vector field $A_\mu$ is specified by
\bea
A_\mu = C \partial_\mu e^{\Phi}.
\eea
We observe that under the transformations (1) 
the tensor ${\cal{F}}_{\mu\nu}$ remains invariant. In addition, 
the 1-form $A=A_\mu dx^\mu$ transforms to $A'=A+ \Lambda$ 
where $\Lambda$ is the following closed 1-form
\bea
\Lambda=-\frac{2C^2 e^\Phi}{C^2 + e^{-2\Phi}}
(F + C H).
\eea
Thus, we are lead to introduce the invariant action 
\bea
S_4 = -\frac{1}{4 g^2_{10}} \int_{\cal{M}} d^{10} x \sqrt{-G}
e^{-3\Phi/2} {\cal{F}}_{\mu\nu}{\cal{F}}^{\mu\nu}_{\rm s},
\eea
where $g_{10}$ is the 10-dimensional Yang-Mills coupling 
constant and ${\cal{F}}^{\mu\nu}_{\rm s}
=G^{\mu\rho}G^{\nu\lambda}{\cal{F}}_{\rho\lambda}
=e^{-\Phi}{\cal{F}}^{\mu\nu}_{\rm E}$.
In the Einstein frame this action takes the form 
\bea
S_4 = -\frac{1}{4 g^2_{10}} \int_{\cal{M}} d^{10} x \sqrt{-G_{\rm E}}
{\cal{F}}_{\mu\nu}{\cal{F}}^{\mu\nu}_{\rm E},
\eea
where ${\cal{F}}^{\mu\nu}_{\rm E}
=G^{\mu\rho}_{\rm E}G^{\nu\lambda}_{\rm E}{\cal{F}}_{\rho\lambda}$.
From this point of view the transformations (1) 
can be interpreted as gauge transformations.

In fact, under the transformation $C \rightarrow C +1$,
which induces $\tau \rightarrow \tau +1$,
the actions $S_1$ and $S_4$ are symmetric. This invariance 
accompanied by the previous symmetry, i.e. invariance under 
$\tau \rightarrow -1/\tau$, 
demonstrate that these two actions possess the full 
$SL(2; \mathbf{Z})$ symmetry.

Now let us indicate the usefulness of the invariant actions.
Since the spacetime is dynamical we should also include the 
kinetic term of the spacetime metric, i.e. the Einstein-Hilbert action
\bea
S_{\rm EH}=\frac{1}{2\kappa^2} \int_{\cal{M}} d^{10} x 
\sqrt{-G_{\rm E}}\;R_{\rm E}\;,
\eea
where the scalar curvature $R_{\rm E}$ is constructed from the 
Einstein metric $G_{{\rm E}\mu\nu}$. This action also has
the $\mathbf{Z}_2$-symmetry. According to our request 
we can select a subset of the invariant actions 
$\{S_i|i=1, 2, 3, 4\}$
as a source of gravity, i.e. the action $S_{\rm EH}$ should 
be added to that subset. Therefore, we receive a gravity 
theory with the $\mathbf{Z}_2$-symmetry. 
In addition, due to the presence of the 1-forms and 9-forms 
field strengths, the actions (14) - (16) are appropriate for 
constructing various solitonic solutions of the D(-1), D7 and NS7-branes. 
More precisely, for realizing the D-instanton in its usual 
form as a solution of the field equations one requires 
the dual, i.e. the magnetic, formulation. Presence of the field 
strength ${\widetilde F_{\rm s}}$ in the actions (14) and (16) 
provide this requirement. 
From the instanton point of view the action $S_1$
reveals an electric picture, in which the instanton would be seen 
as an elementary source coupled to the R-R field $C$.
\section{Dependent invariants}

Making use of the Eqs. (1) we received the invariant 
scalars (3). It is possible to construct many other 
(anti-)invariant scalars from the Eqs. (1). They 
are not independent of the scalars $\sigma_1$ and $\sigma_2$. 
Consequently, the corresponding differential forms and actions
usually are not independent of the previous ones. 
Let us illustrate this fact by the anti-invariant scalar
\bea
C' e^{\Phi'}=-C e^{\Phi}.
\eea
Extracting this equation via the Eqs. (3) ensures us 
that it is not independent. Effect of exterior derivative on 
this equation, or on each of the equations in (3), leads to 
the following independent anti-invariant 1-form
\bea
e^{\Phi'}(F' + C' H')=-e^{\Phi}(F + C H).
\eea
The components of this 1-form inspire an invariant action, i.e.,
\bea
I = -\frac{1}{4\kappa^2} \int_{\cal{M}} d^{10} x \sqrt{-G}
(F_\mu + C H_\mu)(F^\mu_{\rm s} + C H^\mu_{\rm s}).
\eea
We observe that this specific action is not independent, that 
is, $I=S_1+S_2$.

The Hodge dual of the Eq. (28) exhibits an independent anti-invariant 9-form
\bea
e^{\Phi'}({\widetilde F'} + C' {\tilde H'})=-e^{\Phi}({\widetilde F} 
+ C {\tilde H}).
\eea
Wedge product of the Eqs. (28) and (30) defines the invariant 10-form 
\bea
e^{2\Phi'}(F' + C' H')\wedge ({\widetilde F'} + C' {\tilde H'})
=e^{2\Phi}(F + C H)\wedge ({\widetilde F} + C {\tilde H}).
\eea
By summing the Eqs. (11) and (12) and using the identity 
$H \wedge {\widetilde F} = F \wedge {\tilde H}$
this 10-form is reproduced and hence loses its independence. 
The action associated with this 10-form also is 
not independent, i.e. again it is given by $I=I_1+I_2$.

In addition to the indicated invariant scalars, 
the 9-forms ${\widetilde F}$, 
${\widetilde F'}$, ${\tilde H}$ and ${\tilde H'}$ 
also enable us to obtain the following invariant scalars
\bea
e^{2\Phi'}{\widetilde F'}\cdot {\widetilde F'}+{\tilde H'}\cdot{\tilde H'}=
e^{2\Phi}{\widetilde F}\cdot {\widetilde F}+{\tilde H}\cdot{\tilde H},
\eea
\bea
e^{2\Phi'}[2C' {\widetilde F'}\cdot {\tilde H'}+(C'^2-e^{-2\Phi'})
{\tilde H'}\cdot{\tilde H'}]=
e^{2\Phi}[2C {\widetilde F}\cdot {\tilde H}+(C^2-e^{-2\Phi})
{\tilde H}\cdot{\tilde H}],
\eea
\bea
e^{4\Phi'}[(C'^2-e^{-2\Phi'}) {\widetilde F'}\cdot {\widetilde F'}
-2C' e^{-2\Phi'}{\widetilde F'}\cdot{\tilde H'}]=
e^{4\Phi}[(C^2-e^{-2\Phi}) {\widetilde F}\cdot {\widetilde F}
-2C e^{-2\Phi}{\widetilde F}\cdot{\tilde H}],
\eea
where dot product between any two 9-forms ${\widetilde A}$ and 
${\widetilde B}$ is defined by
\bea
{\widetilde A}\cdot {\widetilde B}={\widetilde A}_{\mu_1 \cdot\cdot\cdot \mu_9}
G_{\rm E}^{\mu_1 \nu_1}\cdot\cdot\cdot G_{\rm E}^{\mu_9 \nu_9}
{\widetilde B}_{\nu_1 \cdot\cdot\cdot \nu_9}.
\eea
For our case these 9-forms are Hodge duals of the 1-forms $A$ 
and $B$, and hence we have ${\widetilde A}\cdot {\widetilde B}
=-9!G_{\rm E}^{\mu\nu} A_\mu B_\nu$. Thus, the above invariant scalars are not
independent of the previous ones, i.e. they are equivalent 
to the Eqs. (4) - (6), respectively.

According to the feature of $S_4$ it is obvious that it 
cannot be written in terms of two or three of the actions 
$S_1$, $S_2$ and $S_3$. In addition, 
vanishing of the linear combination $\sum^3_{i=1} a_i S_i$ also 
leads to $a_1 = a_2 = a_3 =0$, which implies that the actions 
$S_1$, $S_2$ and $S_3$ also are linearly independent of each other. 
Therefore, we have four linearly independent actions.
\section{Conclusions}

Two dilaton scalar and axion pseudoscalar enabled 
us to establish various adequate $\mathbf{Z}_2$-invariant 
and anti-invariant quantities such as scalars, 
differential forms and actions. There are some other   
new invariants which are not independent of the 
previous ones. 

Our method inspired the action $S_1$ which is a portion of the 
bosonic part of the type IIB supergravity. Since all the 
actions $S_1$ - $S_4$ have the same derivation method they may 
possess the same pertinence. Therefore, from the point of view of 
the bosonic theories they may have the same desirability,
but each action has its own properties. 
However, from the supersymmetry point of view they have
different desirability.

We observed that the invariant 2-form (i.e. the Eq. (8))
can be interpreted as field strength 
of an Abelian gauge field. Therefore, we were guided to a $U(1)$ 
Yang-Mills theory. The $\mathbf{Z}_2$-transformations exhibit  
the corresponding gauge transformation of this theory. 

Proliferation of the invariant variables demands its own subtlety.
More independent invariants maybe exist.
We illustrated that some of the invariant scalars and 
differential forms are not independent,
and hence the invariant actions which are built from them can be
extracted from the original actions. Note that by utilizing 
the self $\mathbf{Z}_2$-dual forms 
$C_4$ and $F_5= dC_4 +\frac{1}{2}B \wedge dC_2 - \frac{1}{2}C_2 \wedge dB$
and combining them with the discussed invariant forms we can 
construct more invariant variables and actions.

{\bf Some notes}

The goal of this paper is not extension of the type IIB 
supergravity theory. In other words, the invariant actions 
(except $S_1$) are not parts of the type IIB supergravity. This 
implies that they do not need to be $SL(2,R)$ invariant.
However, we observed that the actions $S_1$ and $S_4$ possess 
the $SL(2 , Z)$ symmetry. We only borrowed the dilaton and 
axion fields of the type IIB theory but not more. 
In addition, since our setup is bosonic we neglected the 
supersymmetry. For constructing invariant supersymmetric 
actions one should add the Einstein-Hilbert action to
each of the actions $\{S_i|i=1,2,3,4\}$, then supersymmetrize
the resulted theory. 

{\bf Acknowledgement}:

The author would like to thank M.M. Sheikh-Jabbari for 
useful comments.


\begin{thebibliography}{99}

\bibitem{1}
A. Feinstein and M.A. Vazquez-Mozo, Phys.Lett. {\bf B441} (1998) 40.

\bibitem{2}
E. Konishi and J. Maharana,  Int. J. Mod. Phys. {\bf A25} (2010) 3797.

\bibitem{3}
T.W. Grimm, Phys. Rev. {\bf D77} (2008) 126007.

\bibitem{4}
J.Y. Kim, H.W. Lee and Y.S. Myung, Phys. Lett. {\bf B400} (1997) 32.

\bibitem{5}
C. Vafa, Nucl. Phys. B469 (1996) 403.

\bibitem{6}
K. Peeters, P. Vanhove and A. Westerberg, 
Class. Quant. Grav. {\bf 18} (2001) 843.

\bibitem{7}
G. Policastro and D. Tsimpis, Class. Quant. Grav. {\bf 26} (2009) 125001.

\bibitem{8}
A. Kehagias and H. Partouche, Phys. Lett. {\bf B422} (1998) 109.

\bibitem{9}
A. Kehagias and H. Partouche, Int. J. Mod. Phys. {\bf A13} (1998) 5075.

\bibitem{10}
Z. Bern and D.A. Kosower, Phys. Rev. {\bf D38} (1988) 1888; 
Phys. Rev. Lett. {\bf 66} (1991) 1669; 
Nucl. Phys. {\bf B379} (1992) 451; 
R.R. Metsaev, A.A. Tseytlin, Nucl. Phys. {\bf B298} (1988) 109.

\bibitem{11}
A. Frizzo and L. Magnea, Nucl. Phys. B604 (2001) 92.


\end{thebibliography}
\end{document}